\documentclass[10pt,twocolumn,preprintnumbers,amsmath,amssymb,nofootinbib
,superscriptaddress]{revtex4-1}
\usepackage{graphicx,longtable,mathrsfs,color,array}
\usepackage[hidelinks]{hyperref}
\usepackage[usenames,dvipsnames]{xcolor} 
\usepackage{amssymb,amsmath,mathtools,mathrsfs,slashed} 
\usepackage{epsfig,subfigure,placeins,float} 
\usepackage{booktabs,longtable,ctable,multirow} 
\usepackage{exscale,relsize} 
\usepackage[normalem]{ulem} 
\usepackage{enumerate}
\usepackage{times, mathptmx} 
\usepackage[utf8]{inputenc}
\usepackage{color}
\usepackage{graphicx}
\usepackage{color}
\usepackage{graphicx,graphics}
\usepackage{bm}
\usepackage{tabularx}
\usepackage{rotating}
\usepackage{units}
\allowdisplaybreaks[1]

\graphicspath{{./figures/}}

\newcommand{\md}{\mathrm{d}}

\begin{document}

\title{Theoretical priors in scalar-tensor cosmologies: Thawing quintessence}
\author{Carlos García-García}
\email{carlosgarcia@iff.csic.es}
\affiliation{Instituto de Física Fundamental, Consejo Superior de Investigaciones
Científicas, c/. Serrano 121, E–28006, Madrid, Spain}
\affiliation{ Institut de Ci\`{e}ncies del Cosmos (UB–IEEC), c/. Martí i Franqués 1, E–08028, Barcelona, Spain}
\author{Emilio Bellini}
\affiliation{University of Oxford, Denys Wilkinson Building,
  Keble Road, Oxford, OX1 3RH,  UK}
\author{Pedro G.~Ferreira}
  \affiliation{University of Oxford, Denys Wilkinson Building,
  Keble Road, Oxford, OX1 3RH,  UK}
\author{Dina Traykova}
 \affiliation{University of Oxford, Denys Wilkinson Building,
  Keble Road, Oxford, OX1 3RH,  UK}
\author{Miguel Zumalac\'arregui}
\affiliation{Berkeley Center for Cosmological Physics, University of California at Berkeley and Berkeley Lab, CA94720, USA}
\affiliation{Institut de Physique Th\' eorique, Universit\'e  Paris Saclay 
CEA, CNRS, 91191 Gif-sur-Yvette, France}

\date{Received 7 November 2019; accepted 17 February 2020; published -- 00, 0000}

\begin{abstract}
The late time acceleration of the Universe can be characterized in terms of an extra, time dependent, component of the universe -- dark energy. The simplest proposal for dark energy is a scalar-tensor theory --
quintessence -- which consists of a scalar field, $\phi$, whose dynamics is solely dictated by its
potential, $V(\phi)$. Such a theory can be uniquely characterized by the
equation of state of the scalar field energy momentum-tensor. We find the time
dependence of the equation of state for a broad family of potentials and,
using this information, we propose an analytic
prior distribution for the most commonly used parametrization. We show that this analytic prior can be used to accurately
predict the distribution of observables for the next generation of
cosmological surveys. Including the theoretical priors in the comparison with
observations considerably improves the constraints on the equation of state.\end{abstract}

\maketitle

\section{Introduction.}
\label{intro}

Cosmological observations show compelling evidence that the expansion of the
Universe is accelerating
\cite{Riess:1998cb,Perlmutter:1998np,Abbott:2018wog,Hinshaw:2012aka,Aghanim:2018eyx,Alam:2016hwk}. While, as yet, we have no clear
understanding why, proposals abound involving extra fundamental fields which
may or may not modify gravity on large scales. The consensus, for now, is that the
way forward is to improve observations. Fortunately, the prospects look good:
a slew of powerful surveys mapping out the large scale
structure of the Universe over the coming decade should be able to pin down
the expansion of the Universe to sub-percent
level~\cite{Abell:2009aa,Font-Ribera:2013rwa,Spergel:2013tha,Aghamousa:2016zmz,Hounsell:2017ejq}.

An effective way of describing the accelerated expansion  is to assume that it is driven by an extra,
exotic, source of dark energy, with density $\rho_{\rm DE}$ and pressure,
$P_{\rm DE}$.  One can then define the equation of state of dark energy as
\begin{eqnarray}
w\equiv \frac{P_{\rm DE}}{\rho_{\rm DE}} \label{eos}
\end{eqnarray}
and can, in general, depend on the scale factor of the universe, $a$ or
equivalently, in terms of redshift, $1+z\equiv 1/a$. Characterizing the
acceleration of the universe can then be translated into finding an accurate
description of $w(a)$ \cite{Caldwell:1997ii}. Indeed, current cosmological surveys targeting dark
energy have as their primary goal obtaining accurate measurements of $w(a)$.

It is often useful to use a reduced set of parameters that can accurately
describe complex behaviour. A notable example is that of  primordial
parameters arising from inflation. There one generally has that the evolution
of perturbations at early time is governed by an extra field leading to an
imprint of quasi-scale invariant perturbations on large scales. These
perturbations are incredibly well described by an amplitude, $A_S$, a spectra
index, $n_S$ and a tensor to scalar ratio, $r$ \cite{Lidsey:1995np}. While one
can enlarge this space of parameters, it has been repeatedly shown that these
parameters capture the nature of inflationary perturbations with a high degree
of accuracy. In particular, one can constrain, using the Cosmic Microwave
Background (CMB) for example, the
allowed values of $(n_S,r)$ and then consider the subspace in this plane which
corresponds to particular types of inflationary models. In other words it is
straightforward to identify the physical priors one should apply to $(n_S,\,r)$
for any particular inflationary model.

In this paper we want to emulate what has been done in the case of inflation
and identify an efficient and accurate parametrization for dark energy models.
To do so, we revisit the background parametrization of the scalar field
evolution in scalar-tensor theories of gravity.  We will attempt to construct
an accurate parametrization over a wide range of redshifts and then identify
the correct, physical, range of values for this parametrization. I.e.~we wish
to construct a model for the physical priors, a functional probability
distribution function, {${ P}[{\vec \alpha}]$} where ${\vec \alpha}$ is a set
of time dependent functions which uniquely describes the family of
scalar-tensor theories. Here we focus on the simplest case -- quintessence --
that is described by a time evolving scalar field, $\phi$, with a canonical
kinetic term and a potential, $V(\phi)$
\cite{Ratra:1987rm,Wetterich:1987fm,Ferreira:1997hj,Caldwell:1997ii} (see also \cite{Copeland:2006wr,Tsujikawa:2013fta} for reviews); in this case, the time dependent function that uniquely characterizes a model is $w(a)$. For any
particular choice of parametrization, the problem then reduces to studying the
distribution of the corresponding parameters $w(a)$ for a wide range of models (i.e.~choices of $V(\phi)$) and initial conditions for the scalar field. What we propose to do
here is very much in the spirit of \cite{Marsh:2014xoa} but now we wish to
construct a full model for $P[w(a)]$, an approach which can be extended to full scalar-tensor theories further down the line. Note that our method is complementary to what has been done in \cite{Raveri:2017qvt,Gerardi:2019obr}. While their aim is to reconstruct $w(a)$ 
from observations, we use theoretical assumptions.

To go about constructing a prior for $w(a)$ one needs to adopt a parametrization. A wide
range of proposals have been {put} forward that attempt to fully capture its time
evolution. A favoured parametrization is \cite{Chevallier:2000qy,Linder:2002et}
\begin{eqnarray}
w=w_0+w_a(1-a)\,, \label{wa}
\end{eqnarray}
which can approximate most equations of state close to $a=1$ and is widely
used in current data analysis or in assessing the constraining power of future
surveys \cite{Aghanim:2018eyx,Sprenger:2018tdb}. While this parametrization has proven to be useful in identifying
qualitatively different forms of dark energy, it may not necessarily be accurate
enough to fully characterize the time evolution of dark energy over wide range
of redshifts. For example, for a particular form of dark energy that we will
explore in this paper -- quintessence -- we have that $w>-1$ for all $a$ which
can be easily violated for certain choices of $w_0$ and $w_a$ if one chooses to work
with Eq.~\ref{wa}. We shall see, however, that these concerns are not borne out in practice.

{\it Outline:} In Section~\ref{quint} we describe quintessence, presenting the
underlying mathematics but also laying out the various possible regimes that
have been identified for the equation of state; in Section~\ref{priors} we discuss in some detail how we establish what are physical priors in quintessence models, singling out thawing models as the most natural in terms of intial conditions (barring the existence of tracking behaviour); in Section~\ref{approx} we discuss the approximation scheme we will use, and present how the required
accuracy is evaluated and taking particular care to explain any restrictions;
in Section~\ref{results} we work through a range of models to construct an analytic form for the prior function for the equation of state $w(a)$; in Section~\ref{data} we apply it to a current selection of data, to illustrate the impact the priors have on current constraints on the equation of state; finally in Section~\ref{disc} we discuss our proposal and its caveats.

\section{Quintessence.}
\label{quint}
Consider as a starting point, the Horndeski action \cite{Horndeski:1974wa,Deffayet:2011gz,Kobayashi:2011nu}:
\begin{equation}
S[g_{\mu\nu},\phi]=\int\mathrm{d}^{4}x\,\sqrt{-g}\left[\sum_{i=2}^{5}\frac{1}{8\pi G_{\text{N}}}{\cal L}_{i}[g_{\mu\nu},\phi]\,+\mathcal{L}_{\text{m}}[g_{\mu\nu},\psi_{M}]\right]\,,\label{eq:action}
\end{equation}
where
\begin{eqnarray}
{\cal L}_{2} & = & G_{2}(\phi,\, X)\,,\label{eq:L2}\\
{\cal L}_{3} & = & -G_{3}(\phi,\, X)\Box\phi\,,\label{eq:L3}\\
{\cal L}_{4} & = & G_{4}(\phi,\, X)R+G_{4X}(\phi,\, X)\left[\left(\Box\phi\right)^{2}-\phi_{;\mu\nu}\phi^{;\mu\nu}\right]\,,\label{eq:L4}\\
{\cal L}_{5} & = & G_{5}(\phi,\, X)G_{\mu\nu}\phi^{;\mu\nu} \nonumber\\ & &-\frac{1}{6}G_{5X}(\phi,\, X)\left[\left(\Box\phi\right)^{3}+2{\phi_{;\mu}}^{\nu}{\phi_{;\nu}}^{\alpha}{\phi_{;\alpha}}^{\mu}-3\phi_{;\mu\nu}\phi^{;\mu\nu}\Box\phi\right]\,.\nonumber \\ \label{eq:L5}
\end{eqnarray}
We have that $X=\frac{1}{2}\nabla^\mu \phi\nabla_\mu\phi$ and, for example, $G_{4X}=\partial G_4/\partial_XG_4$. The Horndeski action describes the most general Lorentz invariant, local action in four dimensions, featuring a scalar field on top of the metric and having at most second-order equations of motion on any background. Even if the final aim, beyond the scope of this paper, is to get physical priors for this action in full generality, in this paper we focus on a simpler scenario, quintessence. Quintessence models are a sub-class of Horndeski with this form
\begin{eqnarray}
G_{2}&=&X-V(\phi)\,, \nonumber\\
G_{3}&=&0\,, \nonumber\\
G_{4}&=&\frac{1}{2}M_{P}^{2}\,,\nonumber \\
G_{5}&=&0\,, \nonumber
\end{eqnarray}
where the reduced Planck mass is $M^2_P=1/8\pi G$.
By analogy with perfect fluids, we can define the energy density and pressure of the quintessence field as
\begin{eqnarray}
  \rho_{\rm DE} &=& \frac{1}{2} \dot{\phi}^2 + V(\phi)\,,\\
   \label{eq:rho}
   P_{\rm DE} &=& \frac{1}{2} \dot{\phi}^2 - V(\phi)\,.
   \label{eq:p}
\end{eqnarray}

It is useful to follow \cite{Marsh:2014xoa}  and rewrite the potential as $V = M_P^2 M_H^2 A U(\phi)$, with
\begin{equation}
U(\phi)=f(\phi)+\sum_{n_{\rm min}}^{n_{\rm max}} c_n \xi_n b_n(\phi)\, ,
\label{eq:V}
\end{equation}
where $M_H=100$ Km s$^{-1}$ Mpc$^{-1}$, $A$ is the potential amplitude and
$f(\phi)$ and $b_n(\phi)$ are a choice of basis functions. The leading
  order $f(\phi)$ is perturbatively modified by the sum. The lower limit,
  $n_{\rm min}$, is model specific and the truncation order, $n_{\rm max}$, is
  randomly chosen. Each perturbation order is weighted by a dimensionless
  constant, $c_n$, given by the theory itself and a random contribution from a
  set of dimensionful constants, $\xi_n$. These are drawn from a Normal
  distribution and are meant to increase the different possible $U(\phi)$
  expansions. It is useful to rescale $\phi\rightarrow M_P\phi$ and
  $t\rightarrow M^{-1}_H t$ so that they are dimensionless.

We can now focus on a homogeneous and isotropic universe where the space-time metric is given by $ds^2=-dt^2+a^2(t)(d{\vec r})^2$. The expansion rate can be  derived from the Einstein field equations to give
\begin{eqnarray}
H^2\equiv\left(\frac{\dot a}{a}\right)^2=\frac{1}{3M^2_P}\left(\rho+\rho_{\rm DE}\right)\,, \label{frw}
\end{eqnarray}
where the energy density of matter and radiation is $\rho$. In terms of dimensionless units, the evolution equation for the scalar field is
\begin{eqnarray}
  {\ddot \phi}+3H{\dot \phi}+AU_\phi = 0\,. \label{sfeom}
\end{eqnarray}
These equations completely define the dynamics of quintessence.

Given the definition of $\rho_{\rm DE}$ and $P_{\rm DE}$, one can use Eqs.~\ref{eos} and \ref{wa} to determine the equation of state $w(a)$ and $(w_0,w_a)$. One can envisage two approaches for calculating the latter two parameters. The first one is to consider  a Taylor expansion  of $w(a)$ around $a=1$  (the value of the scale factor today) giving
\begin{eqnarray}
w_0 &=&  \frac{\dot{\phi}^2 - 2 A U(\phi)}{\dot{\phi}^2 + 2 A
        U(\phi)}, \\
\frac{dw}{da} &=&  \frac{2 A}{\dot{a} \rho_{\rm DE}^2} \left [ 3 U(\phi)
        \dot{\phi}^2 H + U_{,\phi} \dot{\phi} \rho(\phi) \right ].  
\end{eqnarray}
where the commas ``$,\phi$'' are partial derivatives with respect to $\phi$ and the
dots are derivatives with respect to the cosmological time. Alternatively one can simply fit the expression in Eq.~\ref{wa} over a range of redshifts. 

\begin{figure}[htb]
  \centering
  \includegraphics[width=8cm]{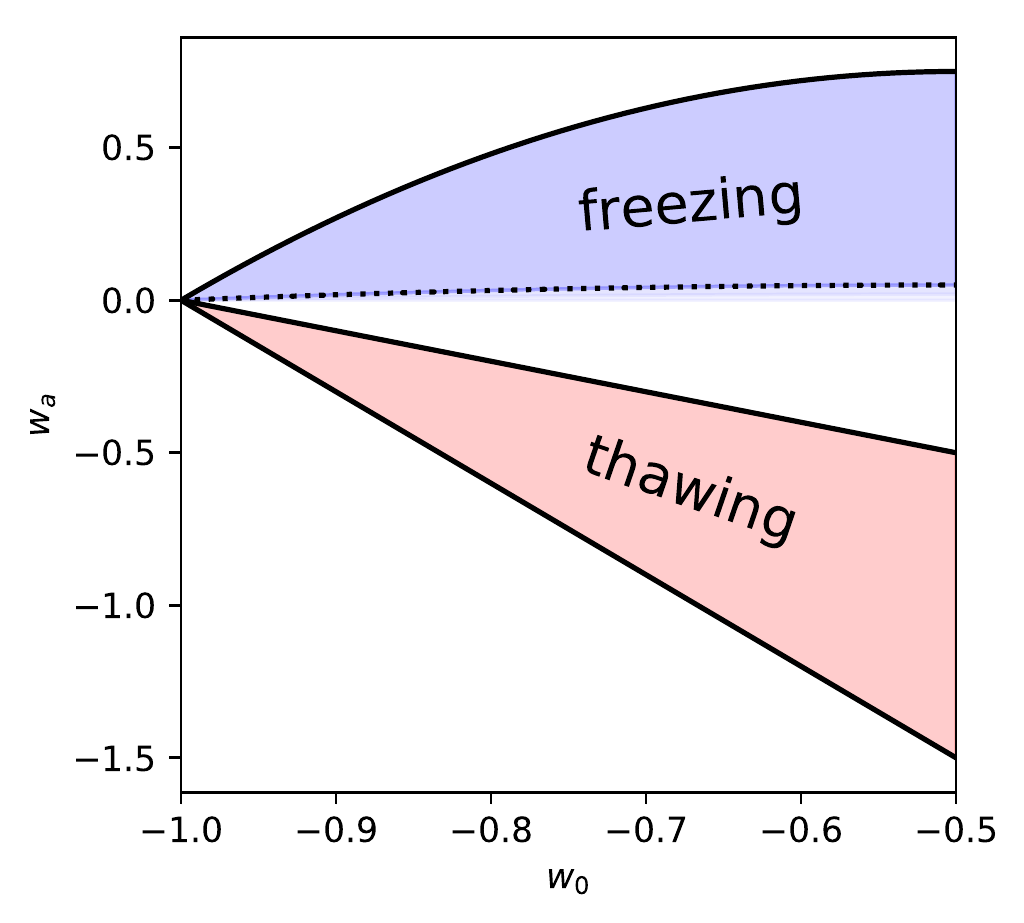}
  \caption{A schematic of the loci of \textit{freezing} and \textit{thawing}
    models on the ($w_0$, $w_a$) plane.}
  \label{fig:schematic}
\end{figure}

It turns out that $(w_0,w_a)$ is a useful 
diagnostic for the types of behaviour that arise. In particular they allow us
to distinguish between two broadly different types of regimes
\cite{Caldwell:2005tm,Barger:2005sb} (see Fig.~\ref{fig:schematic}). In the first regime, the equation of
state can be different from $-1$ in the past but evolves towards $-1$ as the
dynamics of the scalar field freezes with the expansion of the Universe. This
is often called the {\it freezing} or {\it cooling} regime and is 
satisfied (for special choices of initial conditions) by, for example, potentials of the form $U\sim \phi^{-n}$.
In the second regime, the equation of state starts close to $-1$ and
evolves towards larger values as the Universe expands. Often called the {\it
  thawing regime}, it arises, e.g.~in potentials of the form ${\cal
  P}\sim \phi^{n}$.

Whether the quintessence field is in the freezing or thawing regime not only
depends on the form of the potential but also depends heavily on the initial
conditions one is considering. Furthermore, it is possible to choose potentials such that the intermediate
and late time behaviour is an attractor, independent of initial conditions.
Models endowed with this behaviour are known as {\it tracker} models.  In practice, and without tracking, natural initial conditions inevitably lead to thawing behaviour (as we will discuss in Section~\ref{priors}) and, in this paper, we will focus on characterizing these models thoroughly.

\begin{figure}[htb]
  \centering
  \includegraphics[width=\columnwidth]{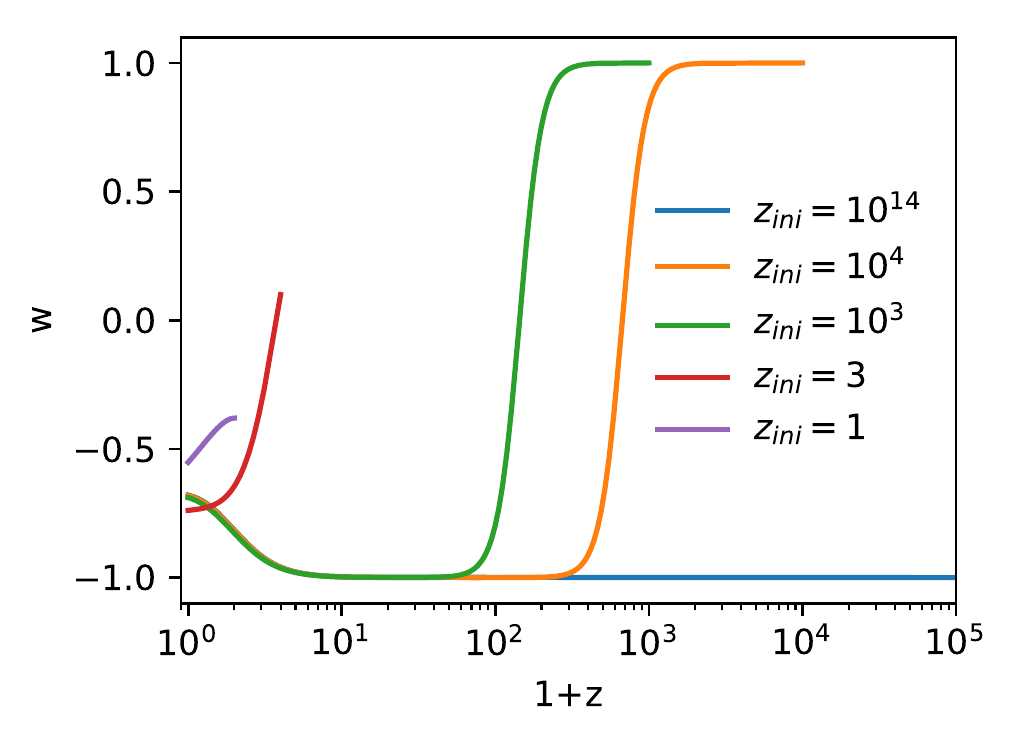}
  \caption{Effect of the initial conditions on the dynamics of the model
    $U \propto \phi^{-2}$ with initial conditions $\phi_i = M_P$ and $\phi'_i
    \equiv \md\phi/\md\tau|_i = H_0$. Just starting the integration close enough $z=0$ the
    initial velocity of the field can survive and yield a freezing behavior.}
  \label{fig:freezing}
\end{figure}

In this paper we will consider a comprehensive suite of dark energy models
(see \cite{Copeland:2006wr,Tsujikawa:2013fta} for a broad range). Building on the choices of
\cite{Marsh:2014xoa} we will explore and characterize the following subclasses
of {\it thawing} potentials given by Eq.~\ref{eq:V} and represented in Table~\ref{tab:models}:\\
\underline {\it Monomial}: $V\sim \phi^N$;  $N$ can be positive, such as the
case of a massive Klein-Gordon field $(N=2)$ or a scale-invariant field
$(N=4)$; or negative. Both cases lead to thawing behavior, provided one solves
the evolution equation from early times. It has been argued that a negative $N$
would lead to freezing behaviour \cite{Caldwell:2005tm,Barger:2005sb}; however, this is only true if the
initial conditions are set at low redshift (see Fig.~\ref{fig:freezing}) or
the kinetic energy is similar to the potential energy at the origin and
$V(\phi_i) \gg V(\phi_0)$ (see discussion in next Section). In
general, starting from {deep} inside the radiation epoch, the Hubble parameter is
large ($\ddot\phi + 3 \dot\phi H \approx 0$) and freezes the field ($\dot\phi
= \dot\phi_i a^{-3}$), which only starts evolving at late times (thawing). In
addition, the early dark energy constraints from CMB and Big-Bang
Nucleosynthesis (BBN) ($\sim 1\%$ \cite{Ade:2013zuv,Ade:2015rim}),
limits the maximum value of $\dot\phi_i$. Recall that, during kination,
$\rho_{DE} \sim \dot\phi^2 = \dot\phi_i^2 a^{-6}$.\\
\underline{\it Effective Field Theory (EFT)}: $V$ is sum of monomials suitably
weighted by the corrected powers of a mass scale consistent with the view that
it arises as a consistent low energy limit of a theory. This series represents
corrections from operators representing high-energy physics. The series
converges if the parameter $\varepsilon_{\rm F}$ is small, required
by the super-Planckian shift symmetry that imposes $\varepsilon_F < 1$.
We will call $n_Q$ to the number of quantum corrections.  \\
\underline{\it Modulus}: $V\sim \epsilon_D^n e^{\lambda_n \phi}$; the vacua potential of theories
with higher order dimensions are usually proportional to exponentials of the
field \cite{Copeland:2006wr}. We consider here weighted sums of them, where
$\epsilon_D$ is the compactification scale. In addition, we parametrize
  $\lambda_n = \alpha (p_D - n)$, so that $\lambda_n$ can be positive or
  negative, depending on the order of the term, and weighted by $\alpha$.\\
\underline{\it Axion}: $V\sim 1+\cos(\epsilon_\mathrm{F} \phi)$ where $\epsilon_\mathrm{F}$ is related to the symmetry breaking scale; these potential arise in theories where the field is a pseudo-nambu Goldstone boson.
Note that we are interested in a regime different of that in which axions contribute to dark matter, in which the field oscillates around a minimum of the potential and the equation of state averages to zero.\\

\begin{table}[tbh]
  \centering
  \begin{tabular}{|c|c|c|c|c|c|c|}
  \hline
  {\bf Model} &  $b_n(\phi)$ & $c_n$ & $n_{\rm min}$ & $f(\phi)$ & $\phi_i$\\
  \hline
  Monomial & 0 & -- & -- & $ \phi^N$ & $[1,7]$\\
  \hline
  EFT & $\phi^n$ & $(\epsilon_\mathrm{F})^n$& $p_E$ & $\sum\limits_{i=2,4}
  \xi_i (\epsilon_{\rm F} \phi)^i$ & $[1, 10]_{\log_{10}}$ \\
  \hline
  Modulus & $e^{\alpha (p_D - n) \phi}$ & $(\epsilon_D)^n$ & 0 & 0 & $[-1, 1]$\\
   \hline
  Axion & $\cos(n \epsilon_\mathrm{F} \phi)$ & $(\epsilon_\mathrm{NP})^{n-1}$ & $2$ & $1+\cos \epsilon_{\rm F}\phi$& $~[-\frac{\pi}{\epsilon_{\rm F}},\frac{\pi}{\epsilon_{\rm F}}]~$ \\
  \hline
  \end{tabular} 
  \caption{Elements of the potential (Eq.~\ref{eq:V}) for the studied
    models. Note that {for EFT, $\phi_i \in [1, 10]_{\log_{10}}$ stands for
    $\log_{10}(\phi_i) \in [0, 1]$.}
}
\label{tab:models}
\end{table}

\section{Establishing Physical Priors}
\label{priors}

Before we attempt to construct an efficient and comprehensive parametrization
of quintessence models it is useful to explore the broad characteristics one
might expect if one focuses on what we determine to be physical priors. We will focus on ($w_0$, $w_a$), very
much along the lines of
\cite{Caldwell:2005tm,Barger:2005sb,Huterer:2006mv,Marsh:2014xoa}. This will
allow us to explain and make explicit our choices that then lead to a
distribution of dynamical behaviours for each individual model.

We can split the choices into a) cosmological parameters that affect that
background evolution  b) parameters in the action (or more specifically, the
potential $V(\phi)$  and c) initial conditions for the scalar field. With
regards to a) we chose to be sufficiently general but not overly so, or the
analysis is impractical. This means that we choose to vary cosmological
parameters that most directly affect the background evolution and which play
off against the scalar field evolution. Specifically, we restrict ourselves
to just $h$ (where the Hubble constant is $H_0=\unit[100\,h]{km\, s^{-1} Mpc^{-1}}$) and
the fractional density of cold dark matter today,
$\Omega_{cdm}$; We choose the ranges $h \in U[0.6, 0.8]$ and  $\Omega_{cdm}
\in U[0.15, 0.34]$. It is important to note that ranges are chosen to be broad
enough so that we do not bias our result, but we do not explore regimes that
are clearly ruled out by current observations (e.g. $H_0 = 0$) \footnote{The
  other parameters are left as the hi\_class default, i.e. Planck 2013
  values. In general, these parameters will have little effect in our
  result, except, perhaps, for the lack of massive neutrinos, whose effect is
  degenerate with DE and could have a small impact when used with data.}.
In particular, $\Omega_{cdm}$ priors might seem too broad in comparison
  with typical constraints. However, they were enlarged to allow for
  compatible results with tomographic weak lensing constraints from
  KiDS-450~\cite{Hildebrandt:2016iqg} or CFHTLenS~\cite{Joudaki:2016mvz},
  which gave weaker constraints on $\Omega_m$.

With regards to parameters in the potential, the choices are more subtle. In
principle one should work within the framework of effective field theory and
only include terms (``operators'') with a suitable dimensionfull weighting and
dimensionless parameters of $O(1)$. Indeed, one of the options makes this
philosophy explicit -- the case of EFT -- and generally we obey this
principle. Nevertheless, if used bluntly, this can severely restrict the range
of behaviours; indeed it will be overwhelming dominated by a cosmological
constant. We take care to allow for more general behaviour to be able to
emerge.

We should note that alternatives are permissible where naturalness still plays
a role (e.g. in the case of monomial potentials that break shift symmetry) or
where other symmetry principles may be at play (as in the case of scale
invariant potentials or axion like potentials). Furthermore, more complex
operators arising from the low energy limit of higher dimensional theories and
their variants (where, for example, factors of $e^{\lambda\phi}$ might emerge)
are also considered. 

Finally, we need to address the initial conditions for the scalar field. We
assume that the initial conditions are established deep in the radiation era
and there we can identify two different regimes. In the first case, the
expansion of the Universe will have damped the evolution such that, in full
generality, ${\dot \phi}=0$. We can, again, use the view point of effective
field theory and assume that the amplitude of the scalar field is of a $O(1)$
in terms of either Planck units or the fiducial mass scale which is used for
dimensionful power counting. Alternatively, one can assume that the scalar
field ``scales'' or ``tracks'' initially, i.e. that its background energy density
follows that of radiation. In that case the initial value and velocity is
uniquely determined by the background cosmology. 

We have chosen these ranges in cosmological and potential parameters as well
as in the initial conditions for the scalar field to span as broad a set of
possibilities as are physically credible, very much in the spirit of
\cite{Marsh:2014xoa}. A list of the ranges is presented in Table~\ref{tab:ranges} 
where the parameters are described in Section~\ref{quint}.
This does mean that we are excluding what we consider unphysical possibilities
such as, for example, starting the scalar field with an arbitrary velocity at
some low redshift -- there is simply no physical mechanism that would put it
there. 
Or, for example, we do not consider hugely disparate sets of parameters in the potential,
as the small parameters would receive large radiative corrections in the absence of a symmetry protecting them.

{We also avoid having too many higher order terms in the potential,
  controlled by $n_{max}$ and $n_Q$, as that would make their contribution negligible. In the same way we don't have either
too low or too high values of the weighting constants $\epsilon_{F,NP,D}$
  that would make the higher order terms either stop contributing or explode.
  Similarly, we choose $\alpha$ and $p_D$ priors for Modulus so that the
  exponent size is controlled, despite allowing for unlikely large values such as
  $e^5$. For the monomial case,  large $N$ values would make the potential too steep
  making the field evolve too fast and yielding unphysical or highly tuned
  results. Finally, the $\xi_n$ constants weigh randomly higher
  order term in the potential, leading us to take their values from a Normal
  distribution.} 

\begin{table}[t]
  \centering
  \begin{tabular}{|c|c|c|c|c|c|c|}
  \hline
  {\bf Parameter} &  {\bf Model} & {\bf Distribution} \\
  \hline
  $A$ & All &  Fixed by $1 = \sum_i \Omega_i$\\
  $\xi_n$ & All &  $\mathcal{N}(0, 1)$ \\
  $N$ & Monomial & $\mathrm{U}_{\mathbb{Z}}[1,7)$ \\
  $n_{\rm max}$ & ~Modulus, Axion ~ & ~$\mathrm{U}_{\mathbb{Z}}[10,20]$~ \\
  $n_{\rm Q}$, $p_E$ & EFT & $\mathrm{U}_{\mathbb{Z}}[5,10]$ \\
  ~$\log_{10}\epsilon_{\rm F,NP,D}$~ & EFT, Modulus, Axion & $\mathrm{U}[-3,-1)$ \\
  $p_D$ & Modulus & $\mathrm{U}_{\mathbb{Z}}[1, 5]$ \\
  $\alpha$ & Modulus & $\mathrm{U}[0, 1]$ \\
  \hline
  \end{tabular}
  \caption{Parameters distributions. $\mathrm{U}[a, b]$ stands for the uniform
    distribution between $a$ and $b$ and $\mathrm{U}_{\mathbb{Z}}$ is the uniform
  distribution in the integers.} 
  \label{tab:ranges}
\end{table}

\begin{figure*}[htb]
  \centering
  \includegraphics[width=\textwidth]{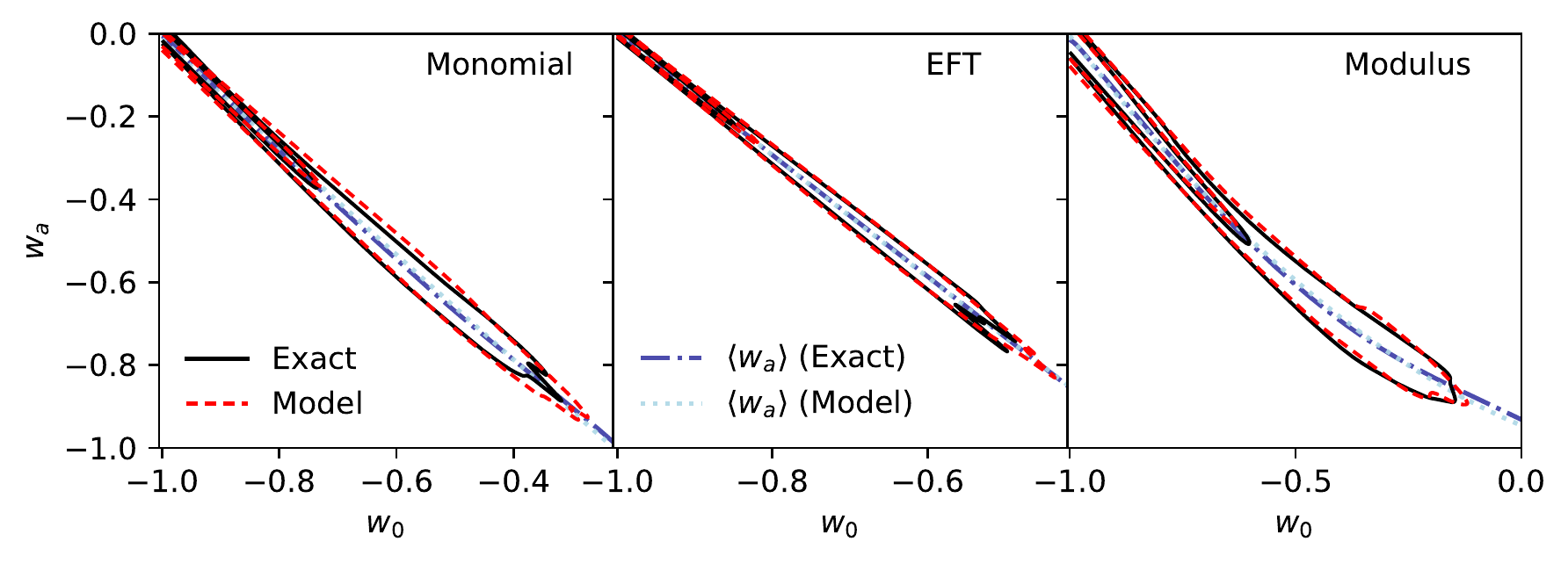}
  \caption{$w_0$-$w_a$ distributions for the Monomial (left panel), EFT (central panel) and Modulus (right panel). All these models have $w_a \neq 0$, while Axion
    can be described just with $w_0 \leq -1 + 10^{-4}$ (95\% C.L.).
    In this Figure we are comparing the prior distributions found for
      $w_0$-$w_a$ fitting the observables (the \textit{Exact}); i.e. minimizing
      Eq.~\ref{eq:chi2}; and the distributions obtained with our analytic
      approximation (\textit{Model}), i.e.  with
      Eq.~\ref{eq:Pw0wa_def}-\ref{eq:Pwa_w0}.}
    \label{fig:w0wa}
\end{figure*}

In Fig.~\ref{fig:w0wa} we can get an idea of the dominant type of
behaviour for the class of models we consider, in terms of the equation of
state. Focusing, for now, on the solid contours, we can see that these models lead exclusively to thawing
models. The only way to change this behaviour (unless the potential tracks the dominant energy density) is to give the scalar field large initial
velocities at late time (see Fig.~\ref{fig:freezing}); these are completely unphysical as they correspond to
unacceptable velocities in the radiation era -- i.e. there is no
physical mechanism for achieving these velocities and hence, such evolution
cannot be, in any way considered part of a set of physical priors. We note
that, in the analysis of \cite{Huterer:2006mv} such unphysical initial
conditions were considered by starting the integration of the background at late redshifts ($z=3$). This prevents the expansion of the universe to dilute the initial conditions and hence the unnatural prevalence of freezing
models in their analysis. We also note that this plot pretty much reflects the
findings of \cite{Marsh:2014xoa}.

\begin{figure}[htb]
  \centering
  \includegraphics[width=\columnwidth]{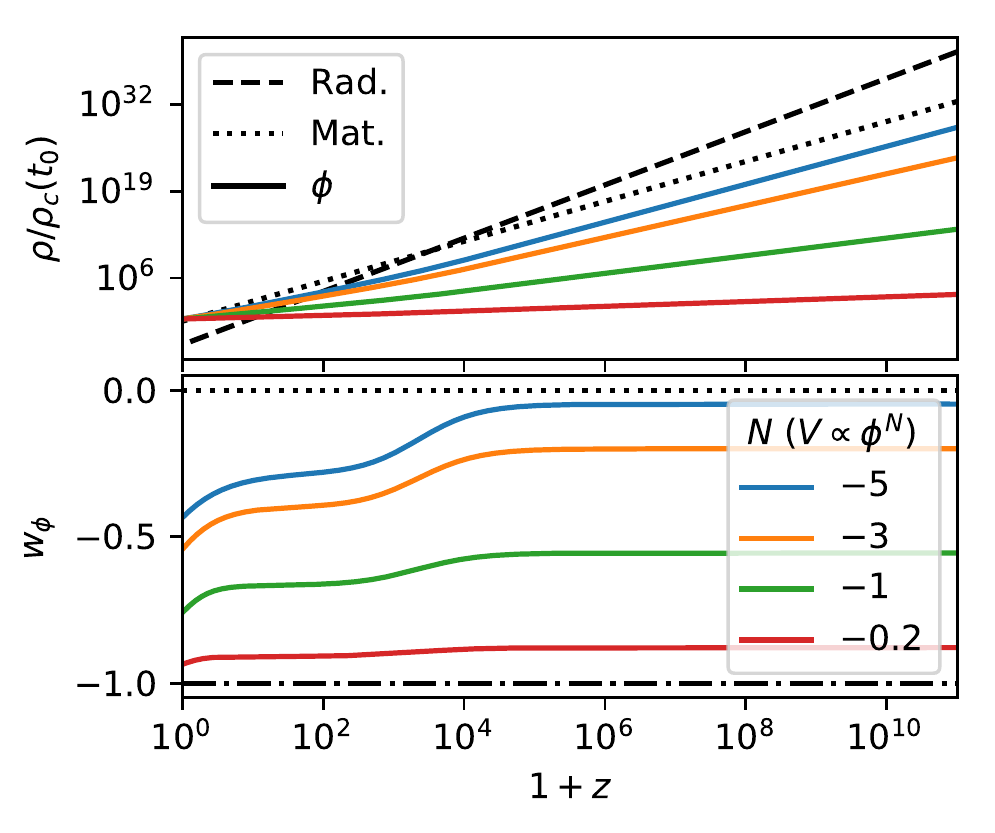}
  \caption{Freezing behavior for inverse power-law potentials (monomial
    quintessence $N<0$). Freezing behavior is possible if the energy density
    of the field is significant at early times $V(\phi_i)\gg V(\phi_0)$ (upper
    panel). However, in order to achieve $w(z=0)\approx -1$ the potential
    needs to be very flat ($N\approx 0$), a requirement that pushes the model
    towards a $\Lambda$ limit (lower panel).}
  \label{fig:freezing_failure}
\end{figure}

Freezing models can be obtained in at least two situations. In \textit{tracker models}, solutions exist in which the quintessence energy density is a fraction of the dominant matter component, e.g. radiation at early times. This early dark energy contributes to the expansion rate throughout cosmic history and is thus subject to constraints from CMB and BBN \cite{Ade:2015rim}.  Generic initial conditions will cause the field to be frozen at $\rho_{\rm DE}\sim V(\phi_{i})$, entering the tracking behavior when $V\sim \rho$. A tracking behaviour at all epochs can be achieved through an exponential potential, although a change on $V_{,\phi}/V$ is necessary for the field to freeze in \cite{Copeland:1997et,Barreiro:1999zs}.  Other freezing models can be obtained with inverse power-law potential (i.e.  monomial with $N<0$) \cite{Ng:2001hs, delaMacorra:1999ff}. In this case the field evolves with a fixed equation of state, rather than track the dominant matter component.  However such a simple potential does not successfully freeze to $w\sim -1$ at low redshift unless $N$ is close to zero, effectively recovering the cosmological constant behaviour, cf. Fig.~\ref{fig:freezing_failure}. In both cases, freezing behaviour requires that the initial conditions for the field are so that $V(\phi_i)\gg V(\phi_0)$ {($\rho_{\rm DE}(\phi_i) \gg \rho_{\rm DE}(\phi_0)$)} -- only thawing behavior can occur if the initial energy density of quintessence is small, as it is the case for the priors used here.  Neither purely exponential or inverse power-law potentials can act as viable freezing models -- we will leave the study of other potentials with freezing behavior for a future study.

\section{Parameterizing quintessence}
\label{approx}

We now proceed to describe the method by which we compress information about
the quintessence component in such a way as to retain as much information as
necessary for a reliable model comparison or likelihood analysis. In this
paper, as it is customary, we choose to work with the equation of state
$w$, which enters into the equations of the observables through the energy
density:
\begin{equation}
  \rho_{\rm DE} = \rho_{\rm DE}^{(0)} \exp\left[-3 \int_0^{log(a)}dlog(a') [1+w(a')]\right]\,.
  \label{eq:X}
\end{equation}
The Hubble rate, $H$, and the angular diameter distance, $D_A$, only depend on the energy densities:
\begin{align*}
  H^2 &\propto \sum_i \rho_i\,,\\
  D_A &= \int_0^z \frac{dz'}{H(z)}\,,
\end{align*}
while the growth rate of structure, $f=d\ln\delta_M/d\ln a$ (where $\delta_M$ is the matter density contrast) also depends (mildly) on $w$:
\begin{align*}
(1+z)\frac{df}{dz} &= \frac{1}{2}\left(1 -
  \sum_i\frac{3w_i\rho_i}{H^2}\right) f + f^2 - \frac{3}{2} \Omega_M(z)\,.
\end{align*}

In choosing a parametrisation, one has to assess how well it approximates the
dynamics it is supposed to represent. To do so, one needs some kind of
measure. For example, it should be good enough that evolution of distinct
classes of models, once parameterised, are easily distinguishable. In practice
this means one needs to construct realisations of this evolution (the equation
of state, $w$, in this case) from a realistic priors for a particular class of
models and check that the parametrisation follows these realisations
faithfully and in such a way that it is clearly distinguishable from another
class of models.

There is also a practical guide that guides the quality of the
parametrisation. Ultimately we will use it to generate mock observables that
will be compared to data. We are looking forward and assuming that the quality
of the data will be such that we will have, at most, percent errors on the
observables. This means that we should aim to have a parametrisation which is
accurate at the sub-percent level but not much more. Another way of stating it
is that the systematic errors arising from our modelling of the dark energy
evolution (through the parametrisation) should be of the same order as (or
marginally smaller than) the statistical and systematic errors arising from
the observations themselves.

We will use the latter criteria as a practical guide in determining the quality of our parametrisation.
The observables we will focus on are the angular diameter
distance, $D_A(z)$, the growth rate, $f(z)$, and the Hubble parameter, over
the range of redshifts which will be covered by the next generation of data
sets. This is key (and a novel aspect) of the approach in this paper.

Our starting point is a finite series expansion of $w$, 
\begin{equation}
  w = \sum\limits_{m=0}^M c_m (1-a)^m\,,
  \label{eq:param}
\end{equation}
Note this parametrization does not break at high order and that higher terms
contribute more at early times ($a \rightarrow 0$). It will be convenient
to denote ${\bf c}=(c_1,c_2,\cdots,c_m)$. 

While it is clear that the larger the number of coefficients (i.e the larger
the $M$), the more accurate the parametrisation, we, naturally, want to come
up with the most economical parametrization. Even though the equation of
state can have a difficult evolution, the results of \cite{Marsh:2014xoa}
and the fact that it will only be relevant when dark energy is not negligible,
lead us to believe that it should, universally, be described by a combination
of a few smooth functions. We shall see that, for thawing models of quintessence, which is the focus of this paper, the parametrization of Eq.~\ref{wa} is sufficient. This means we need to find a probabality distribution function for $(w_0,w_a)$, ${\cal P}[w_0,w_a]$ .

Our approach is as follows \footnote{Our code can be found in
  \url{https://gitlab.com/ardok-m/horndeski-priors }}. For each class of
theories, we generate a large number of possible evolutions for $w$, using the
approach described in Section~\ref{priors}. We work with hi\_class
\cite{Zumalacarregui:2016pph,Bellini:2019syt}, a modified version of the Boltzmann code CLASS
\cite{Blas:2011rf} that incorporates Horndeski theories
\cite{Horndeski:1974wa,Deffayet:2011gz,Kobayashi:2011nu}, to solve the cosmological equations and obtain the
exact observables for each realization.  In addition, the parameters of the model must
be chosen so that we are able to reproduce the phenomenology of the given
model, with mild restrictions based on stability and more fundamental physics.
Given an ensemble of models, parameters are obtained fitting $w_0$ and $w_a$ so
that they minimize the error on the observables at specific redshifts; i.e. minimizing
\begin{equation}
  \chi^2 = \sum_i\frac{(\mathcal{O}({\vec c}) - \mathcal{O}(\phi))_i^2}{\sigma_{\mathcal{O}_i}^2}\,,
  \label{eq:chi2}
\end{equation}
where $\mathcal{O}(\phi)_i$ stands for the exact $D_A$, $H$ and $f$ at the specific redshift
$z_i$, computed solving the field equation and, therefore, depends on the
  specific choice of the model parameters in Table~\ref{tab:models}. On the
  other hand, $\mathcal{O}(\vec c)$ stands for the observables computed using the
  parametrized $w$. As we will see, $\mathcal{O}(\vec c) \equiv
    \mathcal{O}(w_0, w_a)$. Finally, $\sigma_{\mathcal{O}_i}$ is the
  desired precision, which, in all cases, we
choose it to be lower than $1\%$ at low redshift (i.e.~$z<z_{low}=10$)
\cite{Abell:2009aa,Font-Ribera:2013rwa,Aghamousa:2016zmz}, and $0.3\%$ at
recombination for $D_A$ \cite{Joudaki:2017zhq}. Note that, using the parametrization it is possible
that $w$ crosses $w=-1$. To allow for such models that, however, reproduce the
observables accurately, we use the fluid equations with the Parametrized
post-Friedmann (PPF) approximation \cite{Fang:2008sn},
already implemented in CLASS.

Equation~\ref{eq:chi2} neglects the fact that the observables are, in
  general, correlated. However, we can safely disregard this contribution as
  Eq.~\ref{eq:chi2} suffices to achieve the goal precision. This will be seen
  in following Section.

With this machinery in hand, we can study the probability distribution
functions (PDFs) of $(w_0,w_a)$ and from them we are able to generate the
PDFs for the observables, $D_A$, $H$ and $f$.

\section{Results}
\label{results}
We now proceed to construct a distribution function for $w_0$ and $w_a$. 
To begin with, we generate ensembles of cosmologies for each of the models. We follow the procedure we discussed in Section~\ref{priors} to generate $20,000$ realizations for each class of models, which
we have seen is enough to produce stable results. Given the extra accuracy
we need for $D_A(z_{rec})$, we weight the low redshift observables with
$\sigma_{\mathcal{O}_i} = 10^{-3}$ and $\sigma_{D_A(z_{rec})} = 10^{-4}$ for
the angular diameter distance point at recombination in Eq.~\ref{eq:chi2}. Do
not confuse these variables with the maximum allowed errors. They are set
lower to ensure the maximum errors are below our threshold (i.e. 1\% at low
redshift and 0.3\% for $D_A(z_{rec})$).

Revisiting Fig.~\ref{fig:w0wa}, we note that the contours for $(w_0,w_a)$ all have similar characteristics: a long degeneracy direction (given by the dotted line) towards the bottom right while a reasonable tight distribution along the (quasi)-orthogonal direction. {This is characteristic of thawing models, as has been previously observed in \cite{Marsh:2014xoa,Linder:2015zxa}. In the end, the field evolution for thawing models is just a slow (to be physically viable) roll down of the potential and, consequently,  produces similar effects}. We will exploit this structure further down. 

\begin{figure*}[htb]
  \centering
  \includegraphics[width=\textwidth]{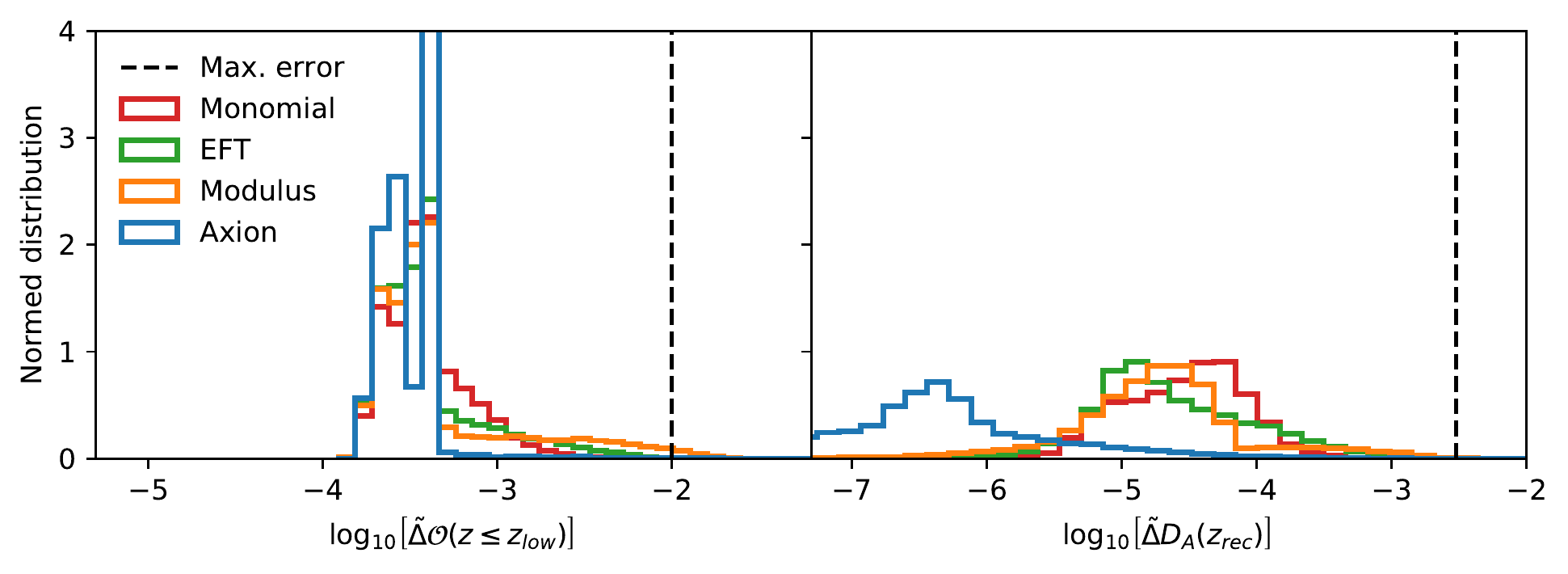}
\caption{Distribution of the maximum relative error of the observables for each Quintessence model compared to its fit in the $(w_0,\,w_a)$ plane, i.e.~$\tilde{\Delta}\mathcal{O}\equiv\left|\mathcal{O}({\rm Fit})/\mathcal{O}({\rm Quintessence})-1\right|$. In the left panel $\mathcal{O}$ stands for any observable among $H$, $f$ or $D_A$, whichever gives the maximum error at low redshift ($z<z_{low}=10$). In the right panel we show the maximum error at recombination for the angular diameter distance $D_A(z_{rec})$.}
  \label{fig:errors}
\end{figure*}

To begin with, as explained before, we check the quality of this parametrization by generating our proxies for the observables, now with $(w_0,w_a)$ and comparing, model by model, with the original full calculation. In Fig.~\ref{fig:errors} we can see that, apart from a very few outliers, most models lie well within our target precision. In other words, using $(w_0,w_a)$ we generate observables which agree with those calculated from the full evolution to better than our threshold. This gives us confidence that describing $w(a)$ in terms of two parameters is good enough for any of the planned surveys.

\begin{figure*}[htb]
  \centering
  \includegraphics[width=\textwidth]{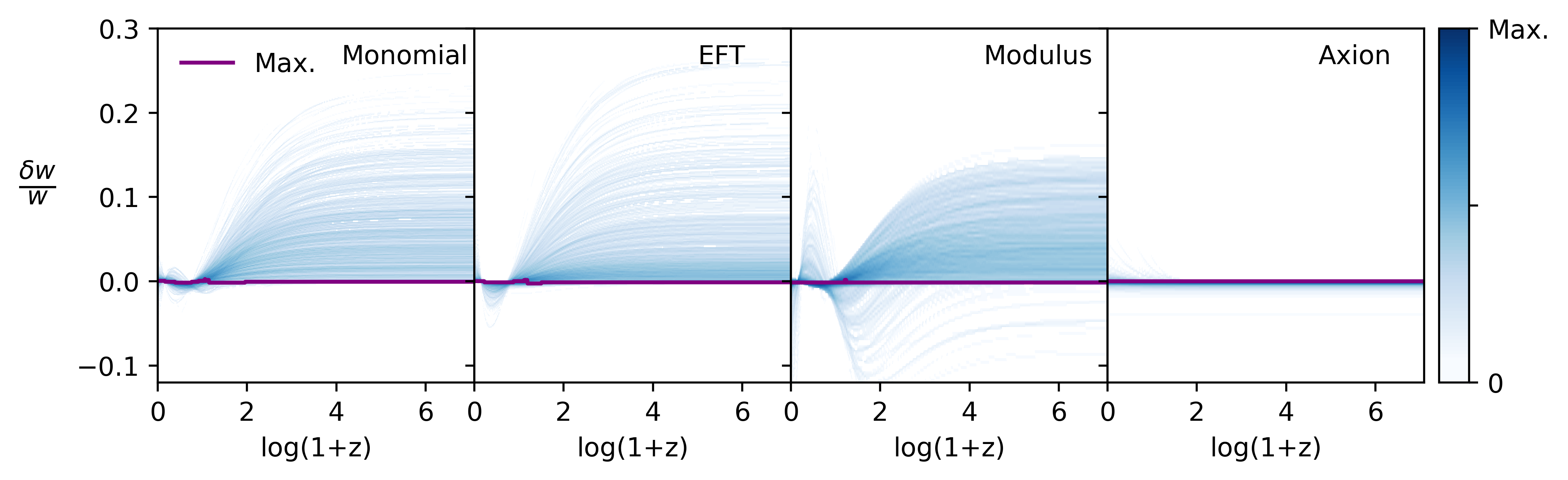}
  \caption{Density plots, one for each Quintessence model, showing the time evolution of the relative difference between $w(a)$ obtained from the original realization and after fitting it in the $(w_0,\,w_a)$ plane. Darker colors indicate the presence of more models with that particular relative difference at that redshift. The purple lines show the maximum density for each model.}
  \label{fig:deltaw}
\end{figure*}

It is interesting to see how well the parametrized $w(a)$ approximates the full one as a function of redshift. In Fig.~\ref{fig:deltaw} we plot the maximum deviation between the two. We can see that for $z>1$, the differences are appreciable. This is, of course, to be expected. The dark energy density at those redshift is sufficiently subdominant and will contribute very little to the observables.

\begin{figure*}[htb]
  \centering
   \includegraphics[width=\textwidth]{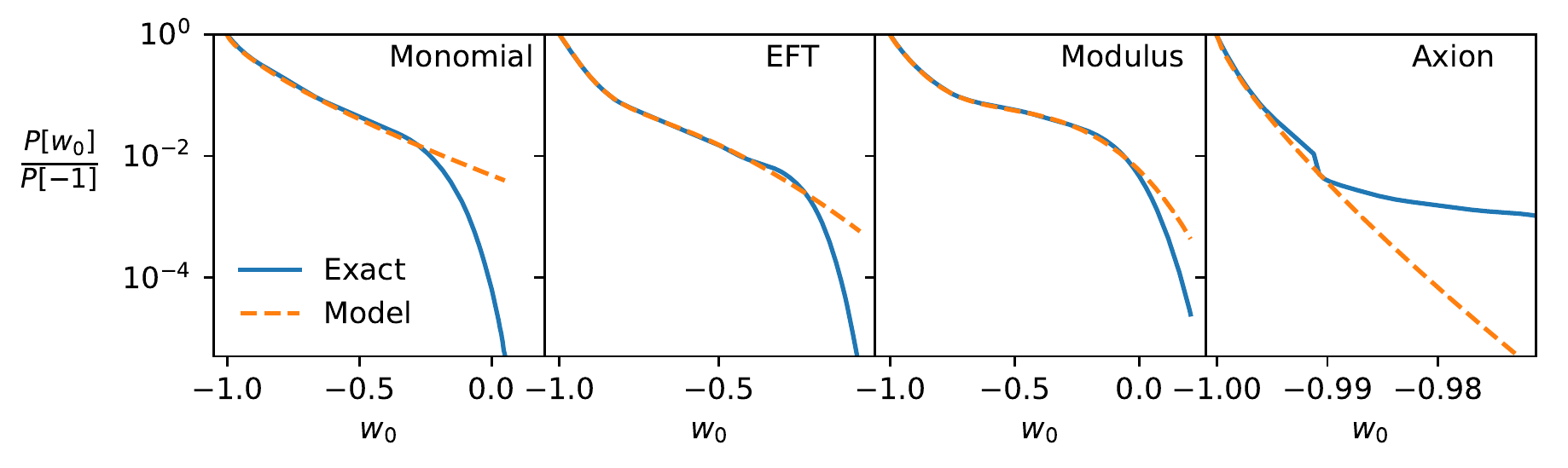}
  \caption{For each model, we show the exact (blue solid lines) probability distribution of $w_0$ normalized to 1 for $w_0=-1$. The approximate analytic model defined in Eq.~\ref{eq:Pw0} is also shown (orange dashed lines). It is possible to appreciate the excellent agreement between the two curves down to probabilities of $10^{-2}$, which includes the region preferred by the data in most models.}
  \label{fig:pw0}
\end{figure*}

With the distributions in hand for $(w_0,w_a)$ we can construct an analytic model. We assume that the distribution can be factorized into 
\begin{eqnarray}
{\cal P}[w_0,w_a]={\cal P}[w_a|w_0]{\cal P}[w_0]\,.
\label{eq:Pw0wa_def}
\end{eqnarray}
We can see in  Fig.~\ref{fig:pw0}, the ${\cal P}[w_0]$ for each model (solid line) is sharply peaked at $w_0=-1$. While an exponential is a reasonable approximation, we find that 
\begin{eqnarray}
{\cal P}[w_0] = A_1 e^{-\left(\frac{w_0}{w_1}\right)^{\alpha_1}} + A_2 e^{-\left(\frac{w_0}{w_2}\right)^{\alpha_2}}
\label{eq:Pw0}
\end{eqnarray}
is more robust and can cover more models. In Fig.~\ref{fig:pw0} we see that approximate form as a dashed line.

\begin{figure*}[htb]
  \centering
  \includegraphics[width=\textwidth]{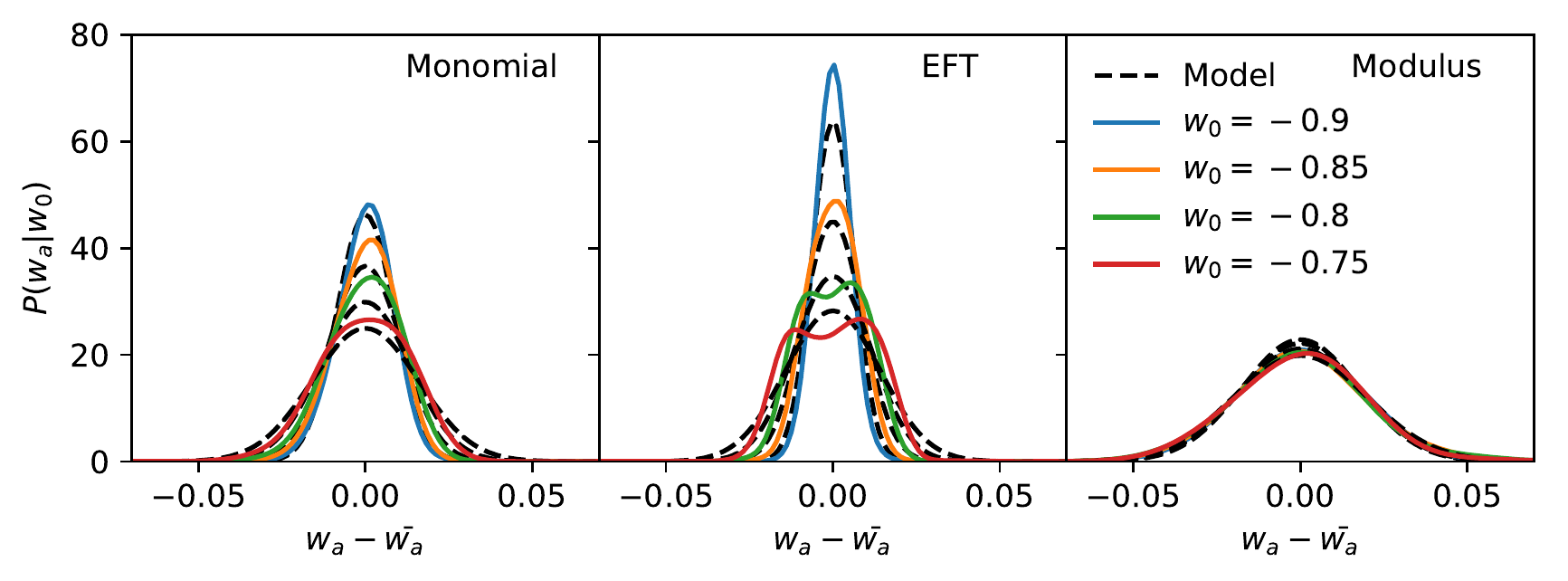}
  \caption{For each model, Axion excluded since it can be parametrized solely in terms of $w_0$, we show the conditional probability distribution of $w_a$ given specific values of $w_0$ (colored solid lines). Dashed black lines represent the same curves obtained using our analytic approximation, Eq.~\ref{eq:Pwa_w0}. The parameters of this equation have been obtained through a global fitting. This explains the discrepancy between the solid and dashed curves, but it does not affect the observables as shown in Fig.~\ref{fig:obs}.}
  \label{fig:deltawa}
\end{figure*}

From the contour plots in Fig.~\ref{fig:w0wa} we can also see that there is a well defined degeneracy direction in $(w_0,w_a)$ (the dotted lines). The degeneracy line, ${\bar w}_a(w_0)$ is well approximated by a quadratic
\begin{eqnarray}
\bar{w}_a(w_0) \simeq \beta_0 + \beta_1 w_0 + \beta_2 w_0^2\,.
\label{eq:mean_wa}
\end{eqnarray} 
We now want to look at the conditional probability, ${\cal P}[w_a|w_0]$. From Fig.~\ref{fig:deltawa} we can see that, for each value of $w_0$, the distribution is peaked around $\bar{w}_a(w_0)$ and well approximated by a Gaussian. We thus use the ansatz
\begin{eqnarray}
{\cal P}[w_a|w_0]=\frac{1}{\sqrt{2\pi\sigma^2(w_0)}} e^{-\left(\frac{w_a - \bar{w_a}(w_0)}{2\sigma(w_0)}\right)^2}\,,
\label{eq:Pwa_w0}
\end{eqnarray}
where the variance is given by
\begin{eqnarray}
\sigma(w_0) \simeq \sigma_0 + \sigma_1 w_0 + \sigma_2 w_0^2\,.
\label{eq:sigma}
\end{eqnarray}
The fits to these parameters for each class of models are given in tables~\ref{tab:Pw0_params} and \ref{tab:Pwa_w0_params}, respectively.
Although they are given with high accuracy, we have found that small
changes do not affect the $w_0$-$w_a$ distribution and, specially, the
distributions of the observables. Nevertheless, we are cautious and use
three significant figures. The study of the minimum required accuracy
goes beyond the scope of this work. 

\begin{table}
  \centering
  \begin{tabular}{|l|c|c|c|c|c|c|}
    \hline
    Model   & $A_1$ & $w_1$ &   $\alpha_1$   & $A_2$ & $w_2$ &   $\alpha_2$\\
    \hline
    Axion   & 737   & 0.775  & $-1.08 \times 10^{-3}$   & 0 & - & - \\
    Monomial   & 8.21  & 0.734  & 0.102                    & 0 & - & - \\
    EFT     & 12.9  & 1.01   & 0.0499        & 1.05  & 1.76  & 0.383 \\
    Modulus  & 8.13  & 0.890  & 0.0712        & 0.535 & 3.95  & 0.803 \\
    \hline
  \end{tabular}
  \caption{Fit parameters for ${\cal P}[w_0]$ (Eq.~\ref{eq:Pw0}) model distributions}.
  \label{tab:Pw0_params}
\end{table}

\begin{table}
  \centering
  \begin{tabular}{|l|c|c|c||c|c|c|}
    \hline
    Model   & $\beta_0$   & $\beta_1$ & $\beta_2$  & $\sigma_0$ & $\sigma_1$ & $\sigma_2$\\
    \hline                                              
    Monomial   & -1.28 & -1.21 & 0.0564               & 0.0769  & 0.108  & 0.0357   \\
    EFT     & -1.43 & -1.35 & 0.0882               & 0.0556  & 0.0577 & 0.00320  \\ 
    Modulus  & -0.946 & -0.460 & 0.481              & 0.0755  & 0.121  & 0.0628   \\
    \hline
  \end{tabular}
  \caption{Fit parameters for ${\cal P}[w_a | w_0]$ (Eq.~\ref{eq:Pwa_w0})
    model distributions. Recall that $\langle w_a \rangle (w_0,\,\beta_i)$
  and $\sigma(w_0,\,\sigma_i)$ are given by Eqs.~\ref{eq:mean_wa} and
  \ref{eq:sigma}, respectively.}.
  \label{tab:Pwa_w0_params}
\end{table}

\begin{figure*}[htb]
  \centering
   \includegraphics[width=0.95\textwidth]{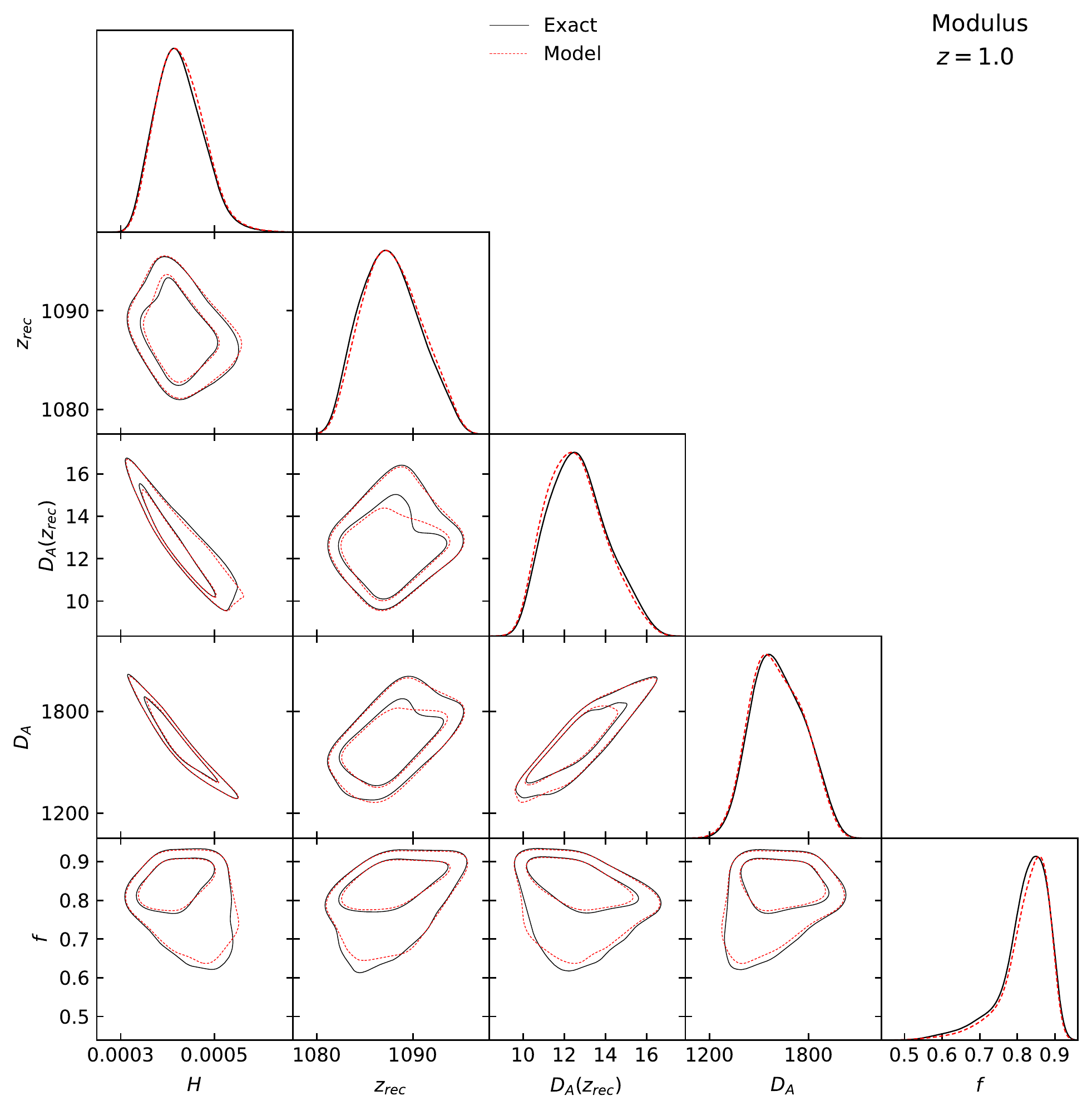}
   \caption{Distributions of the observables obtained integrating the field
     equations of motion (solid black line) compared to those obtained
     sampling from our parametrization. These are the results for modulus at
     redshift $z = 1$, which was the case we found largest differences.
     However, they are small and will have {little} 
     impact when the analytic expressions are used in combination with data.}
  \label{fig:obs}
\end{figure*}

We can now revisit the contour plots, comparing the original ones with those
from our analytical model. In Fig.~\ref{fig:w0wa} we overlay the two sets
and, unsurprisingly, find very close agreement. The final step is now to
sample from the analytic priors to generate the observables and compare
them with those generated by the full scalar field evolution. In Fig.~\ref{fig:obs} 
we overlay the contours for the case with the largest
differences, i.e. for the Modulus model at redshift $z = 1$. We find an
excellent agreement between the two, giving us confidence that the
analytic priors can be used as reasonably accurate representation of the
quintessence space in future cosmological analysis. It must be noted that we
checked the truth of this statement for all models at different redshifts
($z = 0.1,\, 0.5,\, 1.0,\, 2.0,\, 5.0, \mbox{ and } 10$). In
section~\ref{data}, we will study the impact of these analytic priors when
used in combination with actual data and see that they considerably increase 
  the constraining power of the data.

In general, the analytic expression are stable under changes of the
parameters' priors. We checked the stability of the results by approximately
doubling or halving the width of the prior distributions and saw that, even
though (in a very few cases) the actual numbers change, the new distributions can be still
parametrized by Eqs.~\ref{eq:Pw0} and \ref{eq:Pwa_w0}. In particular, we saw
that Monomial and EFT distributions remain unchanged, while the distributions
for Axion and, more significantly, Modulus varied. The exponential potential
of Modulus makes it more sensitive to changes on the field. 

In the following lines we  summarize  some key, individual, features of each model.\\
\underline {\it Monomial:} The two parameters, $w_0$-$w_a$, suffice to
reproduce the observables accurately. In fact, minimizing the $\chi^2$ with
the observables (Eq.~\ref{eq:chi2}) yields equivalent distributions to those
obtained fitting the actual curves (i.e. $w$) at low redshift. The probability
distribution of $w_0$, $P[w_0]$, is correctly reproduced with just one of the
exponentials in Eq.~\ref{eq:Pw0}. The parametrized $w$ goes to lower values
than the original one, meaning that, for a large fraction of the models, one
needs $w < -1$ at early times to effectively recover the model.\\
\underline {\it EFT:} The two parameters, $w_0$-$w_a$, suffice to reproduce the
observables accurately. In contrast with the monomial case, fitting $w$ at low
redshift yields different distributions for the parameters. This is
consequence of faster evolutions of the field close to $z = 0$. In addition,
our prior on $\phi_{i}$ substantially enlarge the $w_0$-$w_a$ distribution in
comparison with that in \cite{Marsh:2014xoa}. This is consequence of having
initial values of the field considerably lower than $1/\epsilon_F$ and forcing
the Friedmann equation to hold today. While in \cite{Marsh:2014xoa} $\phi_i
\in [-1/\epsilon_F, 1/\epsilon_F]$ (with $\log_{10}(\epsilon_F) \in [-3,
-1]$), we imposed a fixed range $\phi_i \in [1, 10]$; the smaller $(\epsilon_F
\phi)^n$ terms that enter into the potential, force the potential to be (substantially)
larger, leading to a large derivative, $V_{,\phi}$. \\
\underline {\it Modulus:} The two parameters, $w_0$-$w_a$, suffice to reproduce
the observables accurately. The more complicated dynamics are reflected in
both the fit performance and the way $w$ is recovered in comparison with the
original: the errors on the observables, although good enough, have longer
tails relative to other models; in addition, in contrast with previous cases,
many models have $w(w_0, w_a) > w(z)$ at early times.\\
\underline {\it Axion:} Just one parameter is needed to reproduce the
observables accurately: $w_0$. In this case, the field remains frozen for
almost all its history and just starts moving very close to the present. As
in the previous case, $w(w_0) > w(z)$ at early times, in a form that
compensates the (small) late time evolution of the field, but does not change
the early universe dynamics.

\section{Comparison with current data}
\label{data}

In this section we present the constraints on the equation of state
from current cosmological data and combine them with the analytical priors of previous section.
Let us briefly go through the method. We  do a Bayesian analysis, sampling through $w_0$ and $w_a$
along with the standard cosmological parameters in a Markov Chain Monte Carlo
(MCMC) with \texttt{MontePython} \cite{Audren:2012wb,Brinckmann:2018cvx} using
the Metropolis-Hastings algorithm \cite{Metropolis:1953am,Hastings:1970aa}.  We have used the Gelman-Rubin
convergence criterion \cite{Gelman:1992zz}, requiring $R-1<0.01$.

We have used the following datasets:\\
\textbf{CMB}: From Planck 2015
\cite{Adam:2015rua,Ade:2015xua,Aghanim:2015xee}, we use the high $l$
temperature autocorrelation (TT) likelihood, in the range $l=30-2508$, along
with the likelihood of the joint autocorrelations of the temperature (TT), the E and
  B polarization modes (EE and BB) and the cross-correlation of the temperature with
  the E-polarization mode (TE) for $l=2-29$ and the lensing likelihood (with temperature and polarisation lensing reconstruction) in the multipole range $l=40-400$.\\
\textbf{BAO}: We use Baryon Acoustic Oscillation (BAO) measurements from 6dFGS survey \cite{Beutler:2011hx}, the Sloan Digital Sky Survey (SDSS) DR7 Main Galaxy Sample (MGS) \cite{Ross:2014qpa}, and BOSS DR12 \cite{Alam:2016hwk}.
The first two measure the expansion history through the redshift-distance and redshift-Hubble relations combined through the relation, $D_{\rm V} = (D_{\rm A}^2(1+z)^2z/H)^{1/3}$, where $D_{\rm V}$ is the angle-averaged distance, $D_{\rm A}$ - the angular diameter distance and $z$ is the redshift
BOSS DR12 constrains both the angular diameter distance, $D_{\rm A}$ and the Hubble parameter, $H$. 
We do not consider the correlation between the BOSS measurements and the
6dF and MGS survey as those surveys cover different patches of the sky and
hence any such correlation would be negligible.\\
\textbf{RSD}: We use Redshift Space Distortions (RSD) measurements derived form from 6dFGS \cite{Beutler:2012px} and BOSS DR12 \cite{Alam:2016hwk}, where for BOSS we use the full covariance between the 3 $f(z)\sigma_8(z)$ measurements at different redshifts and the BAO measurements of $H(z)$ and $D_{\rm A}(z)$.\\
\textbf{SNe Ia}: 
We use the new Pantheon Supernovae (SNe) Ia sample \cite{Scolnic:2017caz}, which 
combines observations from the Pan-STARRS1 Medium Deep Surveyat redshift $0.03
< z < 0.65$ with ones from the SDSS, SuperNova Legacy Survey (SNLS), and various low-redshift and HST
samples. In total 1048 SNe Ia in the redshift range $0.01 < z < 2.3$.

We note that, throughout, we assume that the cross-correlation between the different datasets is negligible.
To obtain the combined constraints from data and theory we implemented the priors as a likelihood module\footnote{Available at:
  \url{https://gitlab.com/ardok-m/horndeski-priors/tree/master/montepython_lkl_quint_thawing_priors}}
in  \texttt{MontePython} and plotted the results with the python package for analysing MCMC samples GetDist \cite{Lewis:2019xzd}.

\begin{figure*}
    \centering
    \includegraphics[width=0.48\textwidth]{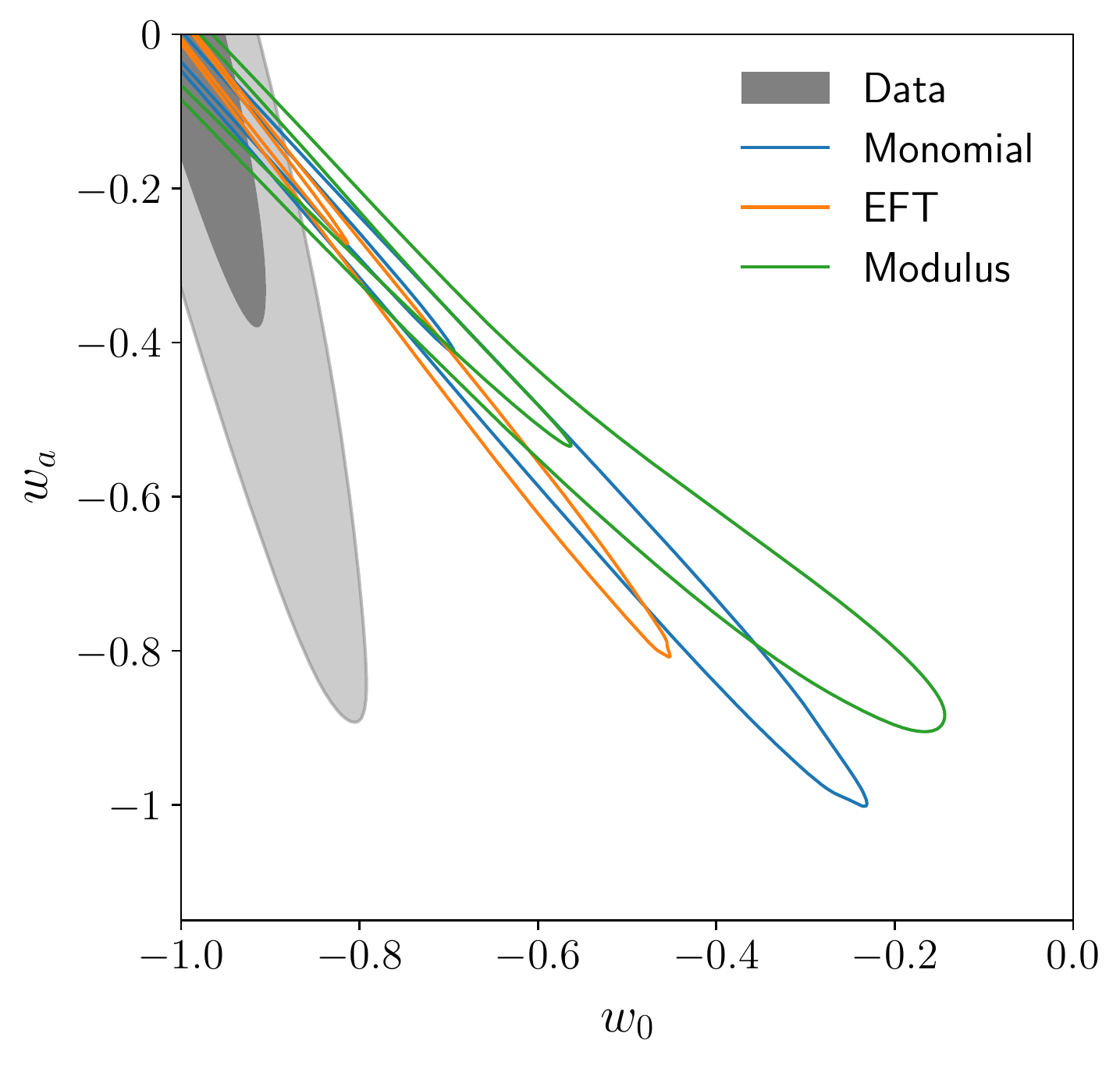}
    ~
    \includegraphics[width=0.448\textwidth]{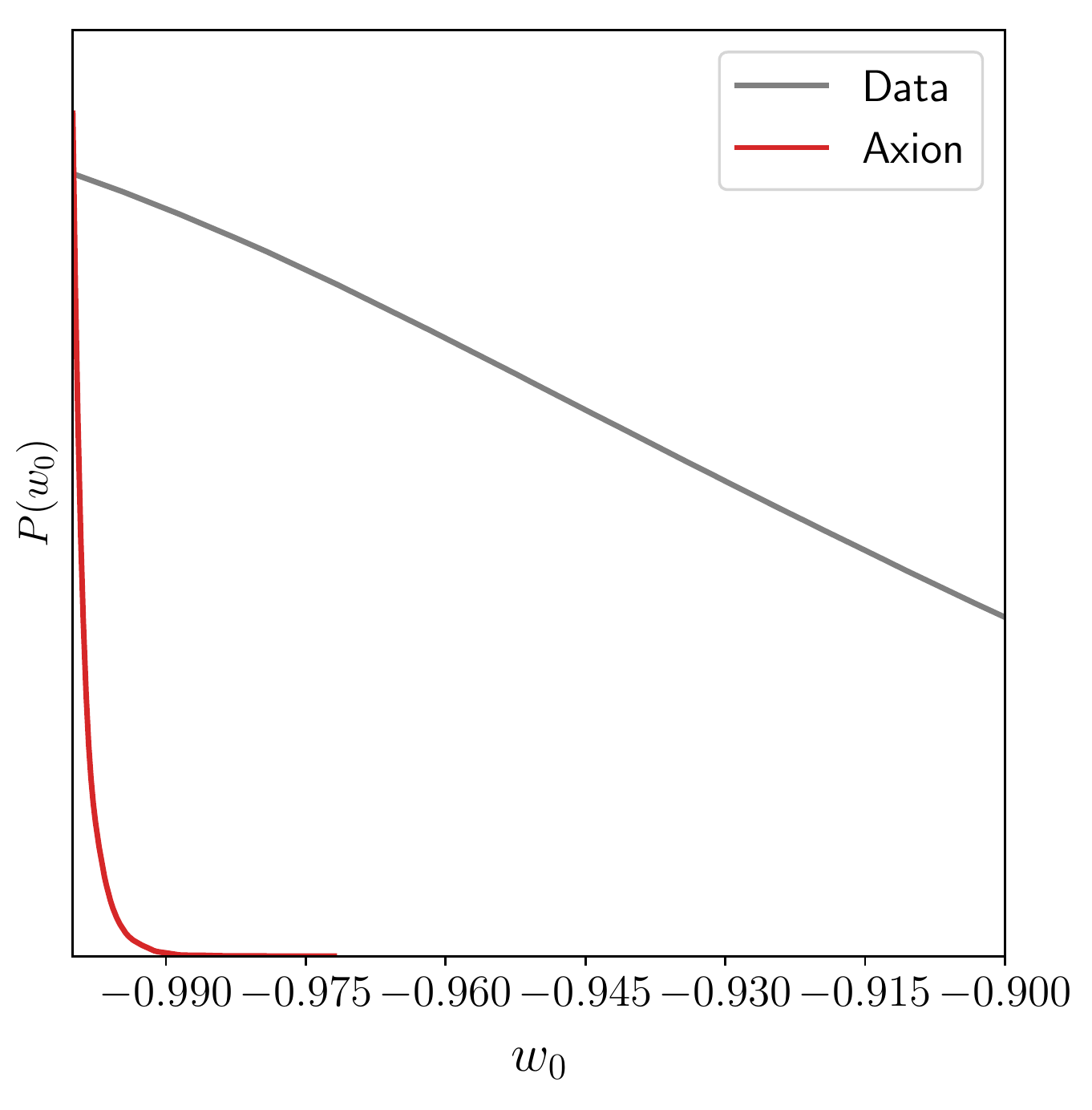}
    \caption{Distribution of $(w_0,w_a)$ for the CMB+BAO+RSD+SN combined
      datasets and the theoretical priors. On the left panel we show
      Monomial, Modulus and EFT and on the right panel Axion quintessence
      priors.}
    \label{fig:data-priors_overlay}
  \end{figure*}

We begin by showing the results of a likelihood analysis without including our model for the theoretical priors in the left hand panel Fig.~\ref{fig:data-priors_overlay}. The likelihood analysis with uniform priors on the equation of state parameters prefers a distinct direction in the $(w_0,w_a)$ plane which, interestingly, is not colinear with the ${\bar w}_a(w_0)$ we derived in the previous section; this is clear from the overlay of the theoretical priors (and was already apparent in \cite{Marsh:2014xoa}). The Axion model, as has been repeatedly mentioned, is a special case, with an equation of state which is very concentrated at $w_0=-1$. For this model, we compare the widths of the posterior from the likelihood analysis with uniform priors on $(w_0,w_a)$ with the theoretical priors studied in this paper; clearly the theoretical prior is much tighter than the posterior. As mentioned above, Fig.~\ref{fig:data-priors_overlay}, to some extent, echoes similar plots used in contour plots for $(n_S, r)$ for constraining the inflationary landscape. There one tends to plot inflationary ``tracks'' for specific models, overlaying the contours from a likelihood analysis assuming uniform priors for $(n_S, r)$.

Even though the two sets of contours overlap, i.e. there is no "tension" between them, we do not expect them to be aligned: very different physical considerations go into generating each set. Our choices were broad and uninformative in order to
          completely explore the phenomenology of the studied models. Furthermore the operational
          assumptions to build the  analytical approximations of the PDF's cannot bias our
          result as they almost perfectly recover the exact $w_0$-$w_a$ distributions.

\begin{figure*}
    \centering
        \includegraphics[width=0.72\textwidth]{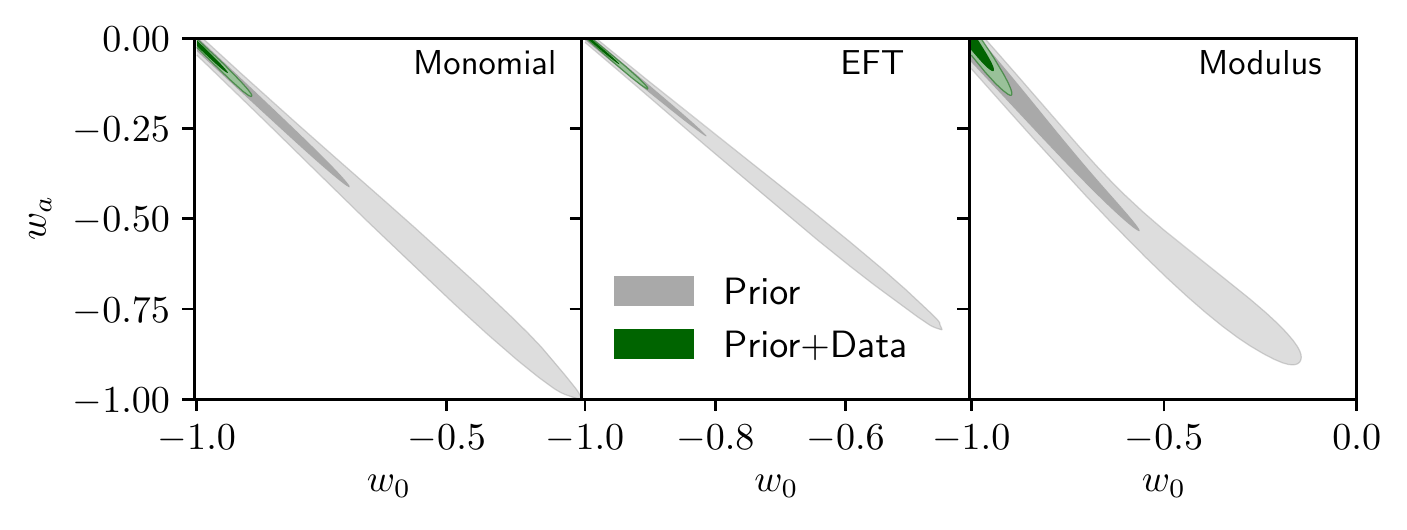}
    \,
        \includegraphics[width=0.25\textwidth]{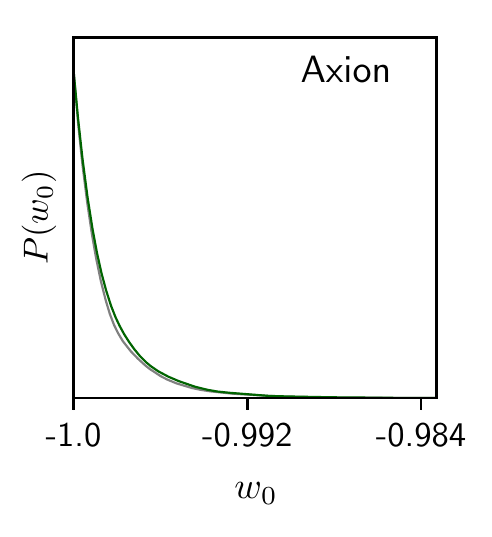}
    \caption{Constraints $(w_0,w_a)$ for the four different quintessence model from the CMB+BAO+RSD+SN combined datasets combined with the priors.}
    \label{fig:combined}
\end{figure*}

If we invert the perspective and plot the theoretical priors individually and then incorporate them in the likelihood analysis of the current data, we see the power of combining the two. In the left hand side of 
Fig.~\ref{fig:combined}, we can see a substantial reduction in the areas of the contours when the data is folded in. We see that the width of the likelihood contours of $(w_0,w_a)$ in the analysis that included the theoretical priors is a factor of few times smaller than for the theoretical priors alone. While one would expect the uncertainty in $(w_0,w_a)$ for the combination of priors and data to be substantially better than for an analysis with uniform priors and data, we can see how the non-colinearity of the prior and data contours works to our advantage and that, even with current data, combining with theoretical priors gives us greatly improved constraints.

To emphasize the power of including correct theoretical priors, in Table~\ref{tab:confidence-lims} we present constraints on $(w_0,w_a)$ for the likelihood analysis with uniform priors and with theoretical priors. Clearly there is a dramatic improvement in constraints. For completeness, again, we single out the case of the Axion model in the right hand panel of Fig.~\ref{fig:combined} to see that in this case the prior is so tightly centred at $w_0=-1$, combining it with data (also centred at $\sim-1$) does not improve the constraints significantly.

\begin{table}
\begin{center}
\begin{tabular}{|p{2cm}|p{3.2cm}|p{2.8cm}|}
\hline
 Data         & $w_0 = -1.012 \pm 0.088$
              & $w_a = 0.05 \pm 0.36$\\
 \hline
 Data + Mono. & $w_0 = -0.966 \pm 0.027$
              & $w_a = -0.058 \pm 0.036$\\
 \hline
 Data + EFT   & $w_0 = -0.968 \pm 0.023$
              & $w_a = -0.040 \pm 0.035$\\
 \hline
 Data + Mod.  & $w_0 = -0.967 \pm 0.025$
              & $w_a = -0.054 \pm 0.037$\\
 \hline
 Data + Axion & $w_0 = -0.9983 \pm 0.0020$
              & $w_a = 0.0$\\
 \hline
\end{tabular}
\caption{Confidence limits of $w_0$ and $w_a$ for the combined data set CMB+BAO+RSD+SN (Data) and the data combined with the theoretical priors for each model.}
\label{tab:confidence-lims}
\end{center}
\end{table}

\section{Discussion}
\label{disc}
In this paper we have proposed a simple, analytic model, for the priors of the equation of state, $w(a)$ of thawing quintessence. We have developed it for four specific thawing models but are confident it could be extended to any thawing model, {given the observed similarities on the $(w_0,w_a)$ space~\cite{Marsh:2014xoa,Linder:2015zxa}}. It can now be used to either demarcate the physically allowed subset of the $(w_0,w_a)$ plane or incorporated into any future MCMC parameter estimation analysis. In Section~\ref{data} we have trialled our proposal with existing data. 

The prior for the equation of state factorizes easily into one probability distribution function for $w_0$ and another for $w_a$ conditional on the value of $w_0$. The model we propose is reasonably robust on the choice of fundamental parameters for each theory (i.e. fundamental constants in the quintessence action and initial values of the scalar field). We mean by this that the priors are unchanged when we vary these fundamental parameters or that the shape of the priors remains invariant (albeit with smaller or larger variances). We believe our choice of fundamental parameters is physical, erring on the conservative side. As a result, any future likelihood analysis is greatly simplified and it is no longer necessary to run the full scalar field dynamics to find the correct distribution of $(w_0,w_a)$. 

One interesting, and important, consequence of having accurate physical priors is that, within the context of thawing models, constraints on $(w_0,w_a)$ are greatly improved. While this is not initially surprising -- the physical priors we have characterized are much more restrictive than the usual uniform priors one uses for $(w_0,w_a)$ -- the fact that the orientation of the prior distribution relative to that of the posterior constructed from the uniform prior with data also plays a key role. In fact, even though the posterior constructed from a uniform prior with data is much broader than the theoretical prior, the fact that they are not co-linear means that combining the two greatly improves the constraints on $(w_0,w_a)$ relative to the prior alone. As the data improves, we except this situation to be further exacerbated.

We have focused on thawing models  but our approach must be extended to other models. Within the scope of quintessence, we have touched on tracker models which lead to freezing behaviour (we do not believe non-tracking, freezing models are physically realistic; {these models need to be tuned so that $V(\phi_i) \gg V(\phi_0)$ and a sufficiently flat potential to correctly approximate to $w\sim -1$, as is exemplified in Fig.~\ref{fig:freezing_failure}}). Tracking models are not amenable to such a simple parametrization. One possible reason is due to the ``quasi-nonanalyticity'' of $w(a)$: in this case, the shape of the equation of state mimics that of rational functions which do not have well behaved Taylor expansions. Nevertheless, there may be alternative ways of parametrizing the equation of state (through, for example Pade approximants) that are amenable to the treatment here.

Naturally, we would like go beyond quintessence and we have tried to couch the task in as general way as possible, as a first step in completely characterizing Horndeski theories. For the case of quintessence, it has been an easy first step given that quintessence models have been extensively studied for over two decades and there is a detailed understanding of how the scalar field evolves in these scenarios. In the case of more general Horndeski theories, pockets have been studied well (such as Galileons or extended Jordan-Brans-Dicke theories) but a complete, generic framework for the scalar field evolution is still lacking. Only once such a framework is in hand can one start discussing physical priors for the parametrized versions of these theories, the ${\cal P}[{\vec \alpha}]$ discussed in the introduction.

\section*{Acknowledgments}
\vspace{-0.2in}

\noindent We thank David Alonso, Eric V. Linder, Alex Malz, Marco
  Raveri and Miguel Aparicio for useful conversations. C.G.G. is supported by AYA2015-67854-P from the Ministry of Industry,
  Science and Innovation of Spain and the FEDER funds, by PGC2018-095157-B-I00 from Ministry of Science,
   Innovation and Universities of Spain and by the Spanish
  grant, partially funded by the ESF, BES-2016-077038. He was also partially
  supported by a Balzan Fellowship while in Oxford. E.B, P.G.F and D.T. are
  supported by European Research Council  Grant No:  693024 and the Beecroft
  Trust. PGF acknowledges support from STFC, the Beecroft Trust and the
  European Research Council. M.Z. is supported by the Marie Sklodowska-Curie
  Global Fellowship Project NLO-CO.  C.G.G. would like to thank New College and
  the Astrophysics department of Oxford University, as well as BBCP,
  UCBerkeley and LBNL, for their hospitality during his short visits there.

\bibliography{./biblio}

\end{document}